# The impact of past pandemics on $CO_2$ emissions and transition to renewable energy

Michal Brzezinski
University of Warsaw, Faculty of Economic Sciences
mbrzezinski@wne.uw.edu.pl

**Abstract**

We estimate the short- to medium term impact of six major past pandemic crises on the CO2 emissions and energy transition to renewable electricity. The results show that the previous pandemics led on average to a 3.4-3.7% fall in the $CO_2$ emissions in the short-run (1-2 years since the start of the pandemic). The effect is present only in the rich countries, as well as in countries with the highest pandemic death toll (where it disappears only after 8 years) and in countries that were hit by the pandemic during economic recessions. We found that the past pandemics increased the share of electricity generated from renewable sources within the five-year horizon by 1.9-2.3 percentage points in the OECD countries and by 3.2-3.9 percentage points in countries experiencing economic recessions. We discuss the implications of our findings in the context of $CO_2$ emissions and the transition to renewable energy in the post-COVID-19 era.

## 1. Introduction

As of April 2021, the WHO data show that the COVID-19 pandemic has resulted in almost 150 million infections and more than 3 million deaths worldwide. The global economy has been hit hard both by the health shock and governments' policy responses aimed at curbing the infections. Recent estimates of the World Bank suggest that the global economy has shrunk by 4.3% in 2020, while the GDP in advanced economies contracted by 5.4% [1]. Although the economy is expected to rebound in 2021, the prospects for overall growth recovery are uncertain. The pandemic has brought significant changes to energy production from fossil fuels, greenhouse gases emissions and the production of energy from renewable sources [2]. Several recent studies estimated the immediate impact of the pandemic and government responses on the reduction of $CO_2$ emissions [3–6]. [7] report that global $CO_2$ emissions are projected to decline by about 7% in 2020. On the other hand, the share of electricity generated from renewable sources has increased in many European countries [8,9]. However, since the global economy will return to growth in 2021 and many of the pandemic-related restrictions are to be lifted we may expect that the observed reduction in $CO_2$ emissions and the rise in the use of renewables could be rather temporary and have little effect on the pace of transition towards a low-carbon economy in the long-run. The long-run impact of the COVID-19 pandemic on the energy sector and climate change depends also on government anti-crisis stimulus packages introduced in 2020-2021 by national governments and intergovernmental institutions like the European Union [7,10]. If those packages contribute to a sizeable increase in clean technology investment as well as discourage the use of fossil fuel-based technologies, the overall effect of COVID-19 and the post-crisis government responses could be accelerating the energy transition. However, at this point the longer run effect of COVID-19 is very uncertain as the pandemic is still an ongoing phenomenon.

In this paper, we attempt to throw some light on the possible short- to medium-term impacts of COVID-19 on $CO_2$ emissions and transition to renewable energy use by looking at the evidence provided by the past pandemics. In particular, we consider the following six major pandemic episodes from the last half of the century: H3N2 Flu (1968), SARS (2003), H1N1 Swine Flu (2009), MERS (2012), Ebola (2014), and Zika (2016). Using the standard macroeconomic method of measuring the effect of exogenous shocks, a local projection approach introduced by [11], we estimate the impact of past pandemics on $CO_2$ emissions and the share of electricity generated from renewables observed over the 5-year horizon after the onset of a pandemic episode.[1] Our baseline data on $CO_2$ emissions comes from the Carbon Dioxide Information Analysis Center (CDIAC) database [12] and cover the period 1751-2017 for 203 nations, while the data on renewable energy is from the U.S. Energy Information Administration's (EIA) international energy statistics [13] and cover up to 201 countries between 1980 and 2018.

To preview our results, we find that the previous pandemics had a significant negative impact on $CO_2$ emissions in the short-run, especially in the OECD countries and in countries that entered the pandemic in bad economic times (recessions). In countries that suffered the most from the pandemics in terms of mortality, the negative impact continued over 8 years. Regarding the transition to renewable energy, our results show no significant effects for the whole sample of countries, but sizeable gains for the OECD countries and countries hit by the pandemics during periods of low economic growth. To provide evidence relevant for the COVID-19, which in general generated a much larger health and economic crisis than the past pandemics that we study, we investigate the heterogeneity of our results with respect to the pandemic severity as measured by the pandemic death toll. The evidence from past pandemics provided in this paper gives a useful benchmark for thinking about the short- and medium-term effects of the COVID-19 pandemic for $CO_2$ emissions and transition to low-carbon technologies. We discuss the implications of our findings for the energy transition in the post-COVID-19 era using early estimates of changes in renewables' use in 2020 and information on green stimulus packages implemented by governments around the world to boost economic growth in an environment-friendly way.

Section 2 of the reminder provides a short description of our data on previous pandemic events, $CO_2$ emissions and the share of electricity generated from renewable sources. We briefly introduce our empirical methodology based on the local projection method in section 3, while our results are presented and discussed in section 4. The last section concludes.

**2. Data**

**2.1. Pandemic events**

We follow [14] in focusing on the six post-war pandemics: H3N2 Flu (1968), SARS (2003), H1N1 Swine Flu (2009), MERS (20212), Ebola (2014), and Zika (2016).[2] Table A1 in the Appendix presents the list of all countries affected during a given pandemic. Our main pandemic-related variable is a dummy marking the timing of the pandemic's outbreak according to the official WHO declarations. However, since the impact of different pandemics varied significantly by country in terms of mortality due to the pandemic we exploit also additional measures of crisis severity. We follow [14] in using three dummy variables capturing the relative health severity of the pandemics. High-severity countries are defined as those that experienced pandemic-related mortality higher than the 70th percentile of the distribution of cross-country mortality due to the pandemic. Medium-severity countries are those in which the mortality was within the range between the 30th and 70th percentile of the distribution, while for low-severity countries the mortality was below the 30th percentile. Our sample covers the 1950-2019 period with 294 country-year observations indicating the start of the pandemic shock.

---

[1] Several papers have recently applied the local projection method to study the effects of past pandemics on various economic outcomes [14,33–35]. The approach has been also used to study the impact of economic and financial crises on $CO_2$ emissions [23,24] and air pollutant emissions [26].

[2] See [14] for a detailed description of these pandemic episodes.

**2.2. $CO_2$ emissions data**

The $CO_2$ data used in this paper come from the Carbon Dioxide Information Analysis Center (CDIAC) database [12] and cover the period 1751-2017 for 203 nations. This data set includes emissions from solid fuel consumption, liquid fuel consumption, gas fuel consumption, cement production, and gas flaring. The CDIAC data are originally expressed in thousand metric tons of carbon. We converted the data to the units of $CO_2$ by multiplying them by 3.667. In our empirical models, we exploit annual country-level emissions measured using the logarithm of $CO_2$ emissions per capita.

In a robustness analysis, we used $CO_2$ emissions data from the Emissions Database for Global Atmospheric Research (EDGAR), which is developed and maintained by the Joint Research Centre of the European Commission [15]. Currently, the EDGAR database offers CO2 emissions data for 202 countries over 1970-2019. The EDGAR $CO_2$ emissions estimates are on average slightly higher than those from the CDIAC as they include also emissions from land-use change (e.g., deforestation) and decomposition of all fossil carbonates. On the other hand, the CDIAC data set starts much earlier allowing us to account also for the 1968 H3N2 Flu pandemic.[3]

**2.3. Share of electricity generated from renewables**

To measure the impact of pandemic shocks on the development of renewable energy, we use a percentage share of renewable energy in total electricity generation.[4] This indicator is the most popular measure of renewable energy deployment [16]. A percentage measure allows making meaningful comparisons across countries and time and indicates the changing role of renewable energy sources in the country's overall electricity mix. Renewable energy data used in the paper are from the U.S. Energy Information Administration's (EIA) international energy statistics on electricity generation) [13]. Renewable energy sources include wind, solar, hydroelectricity, biomass, geothermal, wave, and tidal electricity. The data set covers up to 201 countries between 1980 and 2018. Table A2 in the Appendix presents the descriptive statistics for all variables used.

**3. Empirical methodology**

Pandemic outbreaks are largely exogenous shocks to the economic, health and energy systems. Their impact on the relevant outcomes can be estimated using impulse response functions (IRFs) – a standard macroeconomic tool capturing the dynamic response of a variable to the shock in another variable (see, e.g., [17,18]). While there are several approaches to estimate the IRFs, the local projection (LP) method of Jordà [11] is simple, robust and provides straightforward inference tools [19]. In essence, the LPs are linear regressions of the dependent variable shifted several periods ahead on the current and lagged values of the covariates. In our context, we estimate the following equations:

$$y_{i,t+k} - y_{i,t} = \beta_k D_{i,t} + \gamma_k X_{i,t} + \sum_{j=1}^{m} \delta_j^k \Delta y_{i,t-j} + \alpha_i^k + \tau_t^k + \varepsilon_{i,t}^k, \quad (1)$$

where $y_{i,t}$ is the logarithm of the dependent variable ($CO_2$ emissions or share of electricity generation from renewables) for country $i$ in year $t$, $D_{i,t}$ is a dummy variable indicating that the pandemic shock occurred in the country $i$ in year $t$, $X_{i,t}$ is a vector of pre-determined control variables, $\Delta y_{i,t-j}$ are the lags of the change in the dependent variable (with $m$ set to 2). In some specifications, we replace the pandemic dummy, $D_{i,t}$, with pandemic severity dummies (see section 2.1) to account for the fact that different health crises hit some countries relatively lightly (with no or little excess mortality), while having a very significant impact on other countries. Our control variables include two lags of the pandemic dummy, the GDP per capita and the ratio of trade (sum of exports and imports of goods and services) to the GDP [20]. The lags in the change of the (log of) dependent variable are controlled for since future changes in emissions or electricity generation from renewables could depend on past changes. We also include country fixed effects, $\alpha_i^k$, and time fixed effects, $\tau_t^k$, to control for unobserved, respectively, time-invariant and country-invariant heterogeneity (i.e. global economic or financial shocks).

---

[3] Further details on both databases are discussed by Andrew [36].
[4] Total electricity generation data refer to net generation from utility and non-utility sources from electricity and combined heat and power plants.

In our empirical analysis, we conduct separate estimations of IRFs for $CO_2$ emissions per capita and shares of electricity generation from renewable sources. The estimated $\beta_k$ regression coefficients in (1) represent LP estimates of IRFs to the pandemic shocks happening at time $t$. They measure the percentage change in our dependent variables in countries affected by the pandemics relative to the unaffected countries. We display responses of changes in the dependent variables on figures as the dynamics of $\{\beta_k\}_{k=0}^{5}$ for the time horizons of up to five years after the shock. We also present 95% confidence intervals for our estimates calculated with standard errors clustered at the country level.

The second regression specification that we use allows for testing whether the impact of pandemics is different between advanced and other countries. To this end, we modify the specification (1) by adding a dummy variable representing a given group of countries, $G$, as a control variable and its interaction with the pandemic dummy:[5]

$$y_{i,t+k} - y_{i,t} = \beta_k D_{i,t} + \varphi^k G_{i,t} + \omega^k G_{i,t} D_{i,t} + \gamma_k X_{i,t} + \sum_{j=1}^{m} \delta_j^k \Delta y_{i,t-j} + \alpha_i^k + \tau_t^k + \varepsilon_{i,t}^k. \quad (2)$$

Finally, we use a third model specification to test whether the impact of pandemic shocks on our dependent variables varies with the economic conditions (economic expansions vs recessions) prevailing at the start of the pandemic. For instance, it could be that the impact of pandemic shocks on the transition to renewable energy is more pronounced when the shocks occur in a period of economic recession. Similarly, the fall in $CO_2$ emissions can be higher in a period of economic expansion. To test for this heterogeneity in the impact of pandemics on our dependent variables, we estimate the following models [21–24]:

$$y_{i,t+k} - y_{i,t} = \beta_k^L F(z_{it}) D_{i,t} + \beta_k^L (1 - F(z_{it})) D_{i,t} + \gamma_k X_{i,t} + \sum_{j=1}^{m} \delta_j^k \Delta y_{i,t-j} + \alpha_i^k + \tau_t^k + \varepsilon_{i,t}^k, \quad (3)$$

with

$$F(z_{it}) = \frac{\exp(-\sigma z_{it})}{1+\exp(-\sigma z_{it})}, \sigma > 0$$

where $z_{it}$ is an indicator of the business cycle normalized to have zero mean and a unit variance. Following the literature [21–24], the contemporaneous yearly real GDP growth rate is our choice for $z_{it}$.[6] $F(z_{it})$ is a smooth transition function used to estimate jointly the impact of past pandemics on our dependent variables in economic expansions versus recessions. It can be interpreted as the probability of the economy being in a state of recession. The coefficients $\beta_k^L$ and $\beta_k^L$ capture the effects of the pandemic shocks at the horizon $k$ when the economy, respectively, is in the state of low (negative) economic growth ($F(z_{it}) \approx 1$ when $z$ goes to minus infinity) or high growth ($1 - F(z_{it}) \approx 1$ when $z$ goes to plus infinity). The set of control variables $X$ includes two lags of the following variables: pandemic dummy, the real GDP growth rate, and $F(z_{it})$.

---

[5] The IRFs are then estimated using marginal effects of the pandemic dummy evaluated at the values of the $G$ dummies representing a given group of countries (e.g. OECD or lower-income countries).

[6] The constant $\sigma$ is set to 1.5, but our results are robust to other positive values of $\sigma$.

## 4. Results and discussion

### 4.1. The effect of past pandemics on $CO_2$ emissions

Figure 1 shows IRFs based on LP, estimated using equation (1), for the effect of previous pandemic shocks on $CO_2$ emissions per capita. The estimation sample covers the 1961-2017 period and up to 172 countries. The solid line shows the evolution of emissions in countries affected by the six considered pandemics relative to unaffected countries over five years since the onset of the pandemic shock. The results show that the past pandemics led to a short-term decline in emissions. The pandemics decreased $CO_2$ emissions per capita by 3.4-3.7% within 1-2 years after the start of the shock.[7] After three years, the effect becomes smaller and statistically insignificant. The size of the effect over the 1-2 years after the start of the pandemic is rather moderate. It is approximately equal to 23-25% of the standard deviation of the yearly change in $CO_2$ emissions per capita in our sample. Alternatively, it is quantitatively similar to the 33-36% of the average cumulative change in the $CO_2$ emissions per capita over five consecutive years.

**Figure 1.** The effect of past pandemics on $CO_2$ emissions per capita (CDIAC data, 1961-2017)

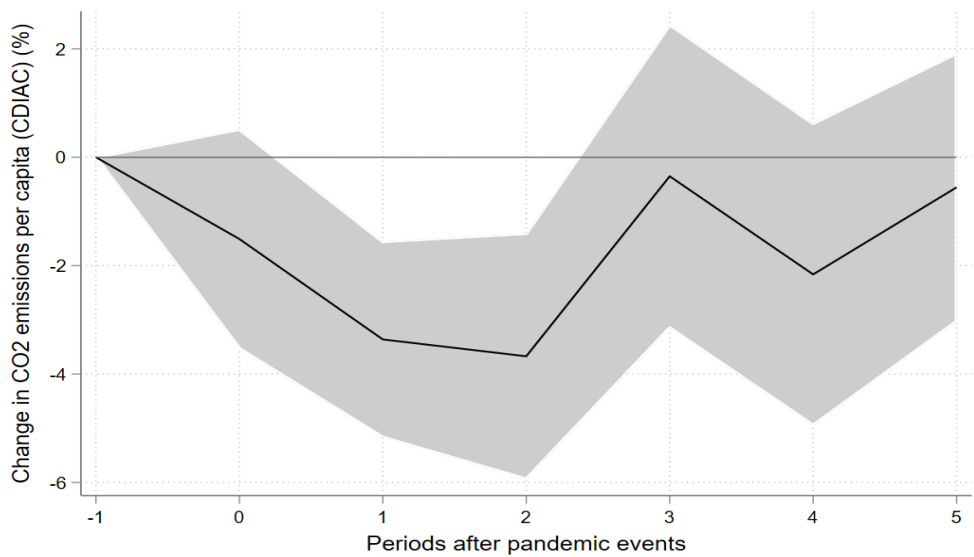

*Note*: Figure shows estimates of IRFs based on LPs (see equation 1). The dependent variable is the change in the logarithm of the $CO_2$ emissions per capita. See section 3 for a description of the model specification. Shaded areas are 95% confidence intervals. The vertical axis shows years ($k$) after the start of the pandemic event with $k = 0$ denoting the year when the pandemic started. The sample covers 172 countries over the 1961-2017 period. Full regression results are available in Table A3 in the Appendix.

---

[7] We obtained very similar estimates using EDGAR data on $CO_2$ emissions. Specifically, the effect of previous pandemics on emissions based on EDGAR data ranges from -2.6% to -3.3% within the four years after the onset of the pandemic episode. Full results based on EDGAR data are available upon request.

Our results can be compared with the early estimates on how $CO_2$ emissions were affected by the COVID-19 pandemic and restrictive measures introduced by governments as a response to the pandemic. The estimates of the 2020 global impact on emissions obtained by Le Quéré et al. [3] are only slightly higher (a 4%-7% decrease) than ours. Similarly, the International Energy Agency (IEA) projected that global $CO_2$ emissions should fall by 8% in 2020 [6]. On the other hand, estimates from near-real-time activity data constructed by the Carbon Monitor initiative (https://carbonmonitor.org/) show that the global emissions declined by 6.2% in 2020 with large heterogeneity between countries [5]. Projections show also that the emissions are expected to rebound globally by 5.8% in 2021 [3]. In China, the reduction in $CO_2$ emissions in 2020 has been estimated to be about 1.6% over the full year, but it is expected that the economic stimulus packages will increase emissions in the next few years [25].

Overall, the existing estimates of the impact of COVID-19 on emissions are quite similar to those obtained in this paper for the case of previous pandemics. While the current shock is probably going to have a somewhat larger immediate impact than the past pandemics, the cumulative decline in emissions associated with COVID-19 is rather temporary and will likely disappear in 2021 or 2022. It is also notable that our estimates are very close to those obtained for the effect of financial crises on $CO_2$ emissions. Financial crises seem to decrease $CO_2$ emissions by 2-4% in a period of a few years after the crises, but the effect seems to disappear in the medium- or longer-term [23,24,26].

Figure 2 provides additional insights into the effects of past pandemic shocks by studying the impacts differentiated by the health shock severity as measured by the relative scale of pandemic-related mortality (see section 2.1). We observe that for medium-severity and low-severity countries the impact of pandemics is mostly insignificant. If anything, it seems to be positive in the fifth year after the onset of a pandemic event in the case of low-severity countries (compared to countries unaffected by the pandemics). On the other hand, we can see a significant and persistent negative impact of the pandemic shocks on $CO_2$ emissions for countries that were hit in the hardest way (with relatively largest pandemic mortality). In these countries, the emissions are lower by 3-4% in the period up to five years compared to the countries unaffected by the pandemics. Figure 2 reveals also that in the fifth year after the start of pandemic events the emissions are 4% lower than in unaffected countries. This effect is sustained over the 8-year horizon after the onset of the pandemics but disappears afterwards.[8] Overall, Figure 2 suggests that the decline in $CO_2$ emissions due to the present crisis should also vary significantly with the size of economic and health cost of COVID-19. In countries that experience the largest GDP falls and the highest human toll due to the pandemic, the emissions should decline more and the effect could last much longer relative to the less affected countries. This seems to create the opportunity for these countries to implement the policies (carbon pricing, public investment in low-carbon technologies, subsidies in green research) that could accelerate their decarbonization pace and reduce the transitional cost to a low-carbon economy. However, such a scenario hinges on the countries' institutional, political and fiscal capacity to introduce bold and radical green efforts during the difficult period of pandemic crisis and recovery.

Figure 3 shows the differential impact of past pandemics on $CO_2$ emissions in rich (OECD) versus lower-income countries.[9] The figure reveals that the negative impact of past pandemics on $CO_2$ emissions is present only in rich countries. If anything, emissions in lower-income countries affected by the previous pandemics are slightly higher after five years compared to the unaffected countries. This observation may reflect the fact that the advanced economies were hit harder by the pandemics or that their governments responded to the past pandemic shocks by introducing more stringent economic restrictions aimed at preventing the spread of diseases.

---
[8] Full estimation results over the 10-year horizon are available upon request.
[9] Lower-income countries are defined as those classified as low-income or lower-middle-income economies according to the World Bank country classification.

**Figure 2.** The effect of past pandemics on $CO_2$ emissions per capita (CDIAC data, 1961-2017) by pandemic severity

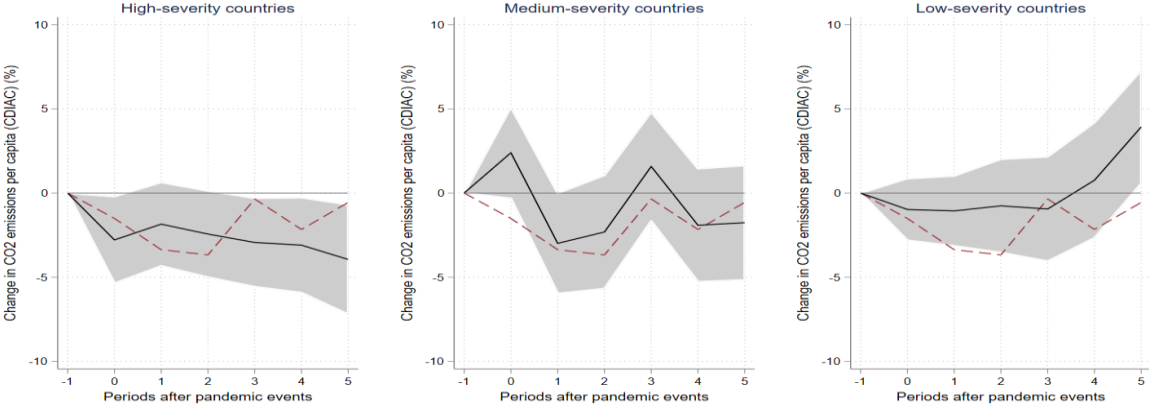

*Note*: Figure shows estimates of IRFs based on LPs (see equation 1). The dependent variable is the change in the logarithm of the $CO_2$ emissions per capita. The solid lines plot estimates for groups of countries defined by pandemic severity, while the dashed line shows the overall effect for all countries (cf. Figure 1). See section 3.1 for a description of the model specification and section 2.1 for the definition of country groups by pandemic severity. Shaded areas are 95% confidence intervals. The vertical axis shows years ($k$) after the start of the pandemic event with $k = 0$ denoting the year when the pandemic started. The full sample covers 172 countries over the 1961-2017 period. Full regression results are available in Tables A4-A6 in the Appendix.

**Figure 3.** The effect of past pandemics on $CO_2$ emissions per capita (CDIAC data, 1961-2017) – the effect in OECD countries and lower-income countries

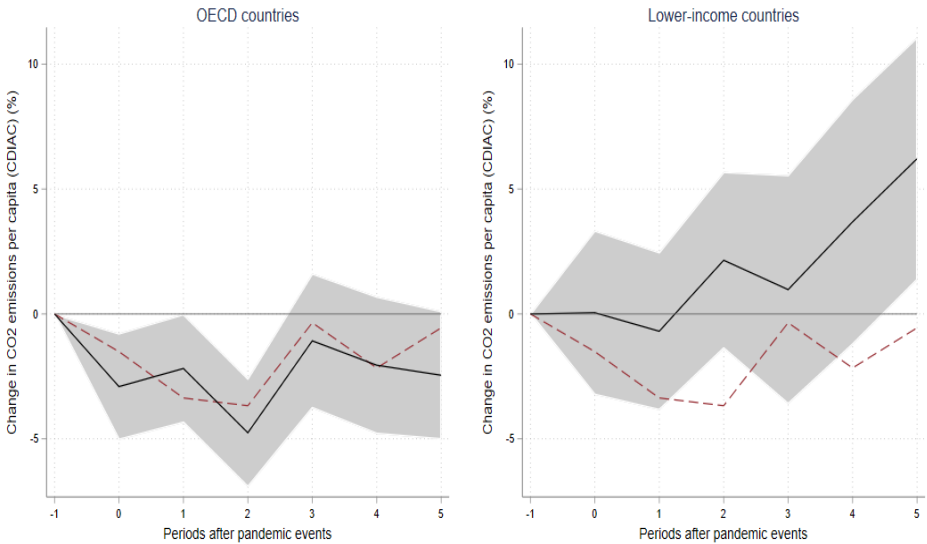

*Note*: Figure shows estimates of IRFs based on LPs (see equation 2). The dependent variable is the change in the logarithm of the $CO_2$ emissions per capita. The solid lines plot marginal effects of the pandemic dummy at the OECD countries (left panel) and lower-income countries (right panel), while the dashed line shows the overall effect for all countries (cf. Figure 1). See section 3 for a description of the model specification. Shaded areas are 95% confidence intervals. The vertical axis shows years ($k$) after the start of the pandemic event with $k = 0$ denoting the year when the pandemic started. The sample covers 172 countries over the 1961-2017 period. Full regression results are available in Tables A7-A8 in the Appendix.

We also analysed how the impact of past pandemics on $CO_2$ emissions depends on the state of the economy during the start of the pandemic. Figure 4 shows that, indeed, the effect is different in periods of high economic growth (economic expansions) and low economic growth (recessions). In expansions, the pandemics have been increasing emissions, but the effect is generally statistically insignificant. On the other hand, economic recessions are associated with significant reductions in $CO_2$ emissions that oscillate around 5% (compared to the countries unaffected by the pandemics) and continue over the 5-year horizon that we study. This result is also consistent with an explanation that pandemic-related reductions in emissions are due to the contraction of economic activity either determined by the health crises themselves or by the government responses that restrict economic activity to contain the pandemic.

**Figure 4.** The effect of past pandemics on $CO_2$ emissions per capita (CDIAC data, 1961-2017) – the role of macroeconomic conditions

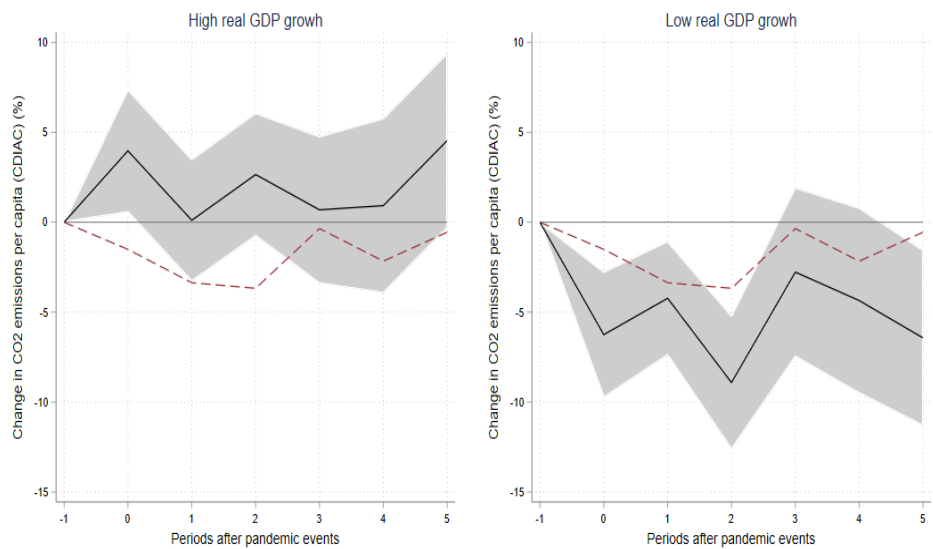

*Note*: Figure shows estimates of IRFs based on LPs (see equation 3). The dependent variable is the change in the logarithm of the $CO_2$ emissions per capita. The solid lines plot estimates for countries experiencing economic expansions (left panel) or economic recessions (right panel), while the dashed line shows the overall effect for all countries (cf. Figure 1). See section 3 for a description of the model specification. Shaded areas are 95% confidence intervals. The vertical axis shows years ($k$) after the start of the pandemic event with $k = 0$ denoting the year when the pandemic started. The sample covers 172 countries over the 1961-2017 period. Full regression results are available in Table A9 in the Appendix.

## 4.2. Previous pandemics and transition to renewable energy

Our baseline estimates of the impact of past pandemic shocks on the share of electricity generated from renewable sources are displayed in Figure 5. The average effect in the full sample is positive but statistically insignificant throughout the five-year horizon after the onset of the pandemic. In Figure 6, we look at the effect differentiated by the pandemic severity. For the countries hit hardest by the pandemics, the pandemic-induced transition to renewable energy use reaches 9.5% (compared to the unaffected countries) in the third year since the onset of the pandemic and becomes statistically significant at the 10% level. However, soon after that, the effect becomes smaller and insignificant. On the other hand, we do not find a positive impact of pandemics in the medium-severity group of countries, while in the case of low-severity ones we observe a negative transitory impact in the fourth year after the start of the pandemics. Overall, we find some evidence that in the short run the countries hit hard by the past pandemics have increased their share of electricity generated from renewable sources. However, this effect does not appear in countries that experienced relatively moderate or small health crises.

**Figure 5.** The effect of past pandemics on the share of electricity from renewables (EIA data, 1980-2018)

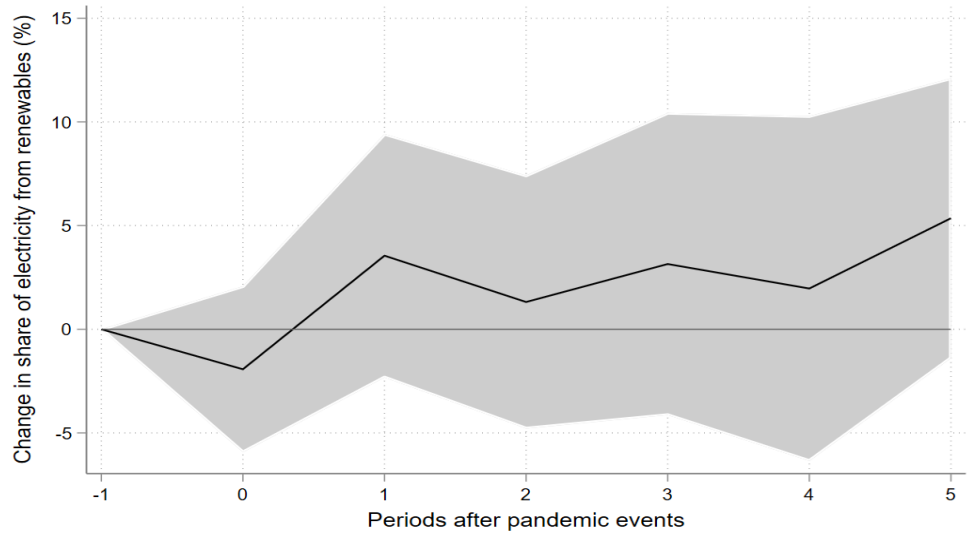

*Note*: Figure shows estimates of IRFs based on LPs (see equation 1). The dependent variable is the change in the logarithm of the share of electricity generated from renewable sources. See section 3 for a description of the model specification. Shaded areas are 95% confidence intervals. The vertical axis shows years ($k$) after the start of the pandemic event with $k = 0$ denoting the year when the pandemic started. The sample covers 174 countries over the 1980-2018 period. Full regression results are available in Table A10 in the Appendix.

We also check whether the impact of past pandemics on the transition to clean energy was different in the OECD versus lower-income countries (Figure 7). The effect for the OECD countries is larger than that for all countries and becomes statistically significant in the first and the fifth year after the onset of the pandemic. In those years, the previous pandemics increased the share of electricity generated from renewable sources in the OECD countries by 6-7% relative to the countries unaffected by the pandemics. Since the average share of electricity generated from clean sources in the OECD countries in our sample is 32.3%, the past pandemics increased the share on average by 1.9-2.3 percentage points within the five-year horizon. The magnitude of the effect is therefore quite substantial and suggests that the rich countries took the opportunity created by the pandemic crises to speed up the transition to clean energy use, at least in the short run.[10] Unfortunately, we do not observe significant effects of pandemics on the transition to clean energy in lower-income countries. Finally, Figure 8 looks at the role of business cycle conditions at the start of the pandemic episodes. It shows that during the periods of economic expansion the transition to renewable energy seems to be in general negligible. On the other hand, in bad economic times, we observe the consistently positive impact of the pandemics on the increase in the share of electricity generated from renewable sources. In the first and the fifth year since the onset of previous pandemic events, the effects are statistically significant and vary in the range from 10% to 12%.[11]

---

[10] The effects for the OECD countries in the longer horizon of 10 years are also positive, but less precisely estimated and statistically insignificant (details are available upon request).
[11] For an average country, this translates into 3.2-3.9 percentage points increase of the electricity share generated from renewables.

**Figure 6.** The effect of past pandemics on the share of electricity from renewables (EIA data, 1980-2018) by pandemic severity

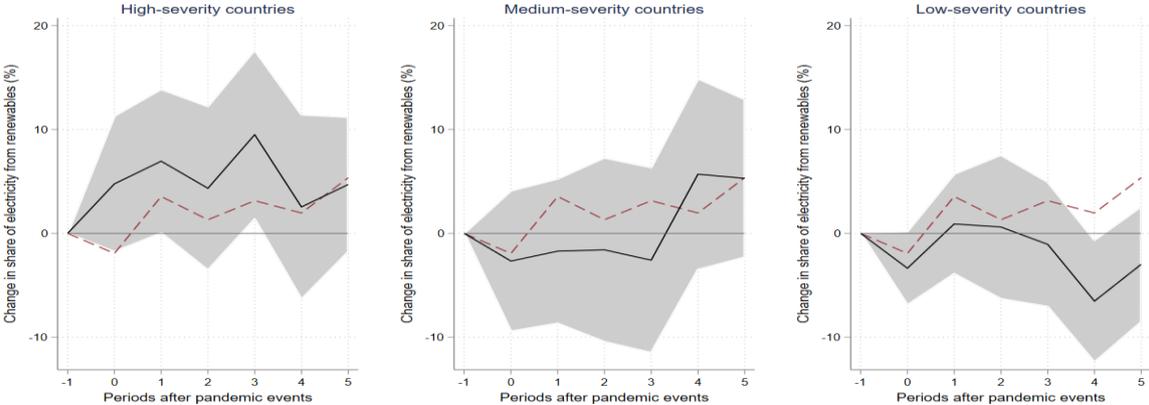

*Note*: Figure shows estimates of IRFs based on LPs (see equation 1). The dependent variable is the change in the logarithm of the share of electricity generated from renewable sources. The solid lines plot estimates for groups of countries defined by pandemic severity, while the dashed line shows the overall effect for all countries (cf. Figure 5). See section 3.1 for a description of the model specification and section 3.2.1 for the definition of country groups by pandemic severity. Shaded areas are 95% confidence intervals. The vertical axis shows years ($k$) after the start of the pandemic event with $k = 0$ denoting the year when the pandemic started. The sample covers 174 countries over the 1980-2018 period. Full regression results are available in Tables A11-A13 in the Appendix.

**Figure 7.** The effect of past pandemics on the share of electricity from renewables (EIA data, 1980-2018) – the effect in OECD countries and lower-income countries

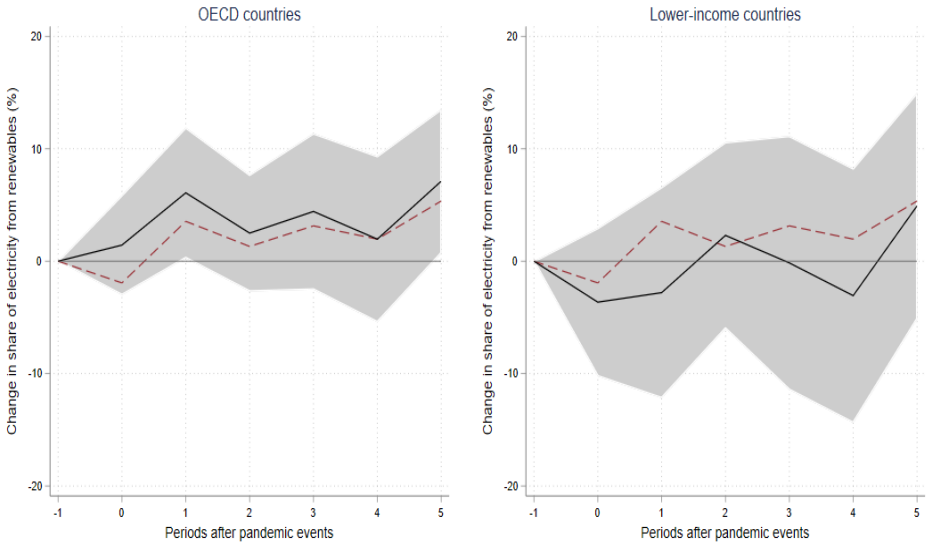

*Note*: Figure shows estimates of IRFs based on LPs (see equation 2). The dependent variable is the change in the logarithm of the share of electricity generated from renewable sources. The solid lines plot marginal effects of the pandemic dummy at the OECD countries (left panel) and lower-income countries (right panel), while the dashed line shows the overall effect for all countries (cf. Figure 5). See section 3 for a description of the model specification. Shaded areas are 95% confidence intervals. The vertical axis shows years ($k$) after the start of the pandemic event with $k = 0$ denoting the year when the pandemic started. The sample covers 174 countries over the 1980-2018 period. Full regression results are available in Table A14-A15 in the Appendix.

**Figure 8.** The effect of past pandemics on the share of electricity from renewables (EIA data, 1980-2018) – the role of macroeconomic conditions

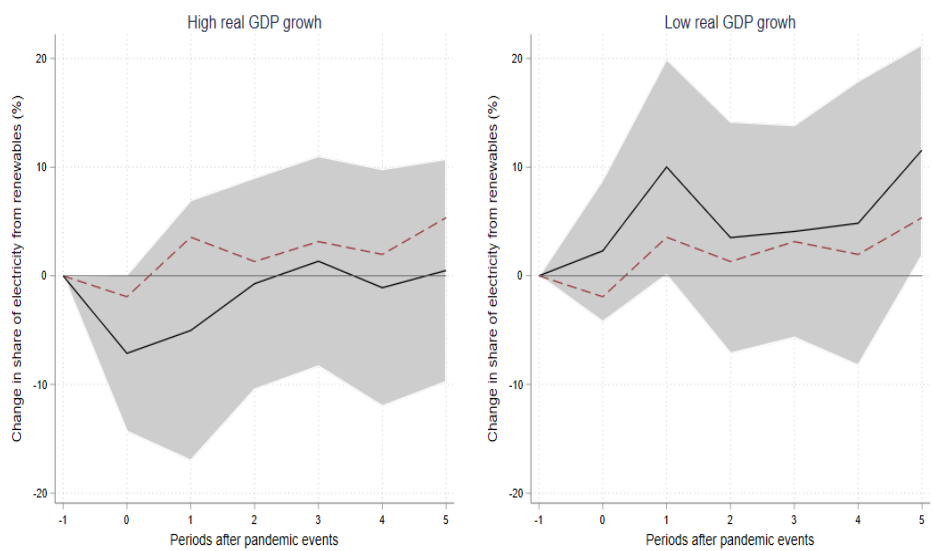

*Note*: Figure shows estimates of IRFs based on LPs (see equation 3). The dependent variable is the change in the logarithm of the share of electricity generated from renewable sources. The solid lines plot estimates for countries experiencing economic expansions (left panel) or economic recessions (right panel), while the dashed line shows the overall effect for all countries (cf. Figure 5). See section 3 for a description of the model specification. Shaded areas are 95% confidence intervals. The vertical axis shows years ($k$) after the start of the pandemic event with $k = 0$ denoting the year when the pandemic started. The sample covers 174 countries over the 1980-2018 period. Full regression results are available in Table A16 in the Appendix.

Overall, the evidence in Figures 6-8 is consistent and implies that the effect of previous pandemics on the transition to renewable energy use is heterogenous across groups of countries. In particular, we found significant effects in the countries hit hard by the pandemics, in the OECD countries, and in countries that experienced low economic growth at the start of the pandemic episodes. The reverse is true in countries that experienced pandemics of moderate or low severity, in lower-income countries, and in countries hit during good economic times. With our data, we are unable to test for the mechanisms that facilitated the energy transition in the groups of countries that we identified. However, it seems obvious that, in general, the rich countries have a higher capacity for the green transition in terms of the institutional framework, political attitudes of citizens, infrastructure and fiscal space. More revealingly, our analysis shows that the opportunity for energy transition arises during particularly hard health crises and bad economic times. Under these circumstances, the overall electricity demand is often reduced which facilitates energy transition, especially when a country is already producing a large share of its electricity from renewables and when the crisis occurs during favourable weather conditions.

Although it is not straightforward to adapt lessons from past pandemic episodes for the current one, we nevertheless notice that there is already some evidence suggesting that the COVID-19 pandemic may be an instigator for a successful energy transition, at least in the advanced countries. It has been shown that in Germany the share of electricity generated from renewables grew above 55% in the first half of 2020 (as compared to 47% in 2019) without any changes to the security of electricity supply or grid stability [8]**.** Electricity generation from intermittent renewables increased as well in many other European countries [9]. However, to ensure the sustainability and resilience of energy systems that rely more heavily on renewable sources countries must strengthen their system flexibility to match the energy demand with relatively more variable electricity supply from renewables [27]. On the other hand, as argued in [10] the worldwide economic recession caused by COVID-19 could lead to a reduction of investment in clean energy technologies and outweigh in the longer run any gains associated with the early increased use of renewables. Ultimately, the overall effect of COVID-19 on the energy transition will depend on government responses [7]. While it is too early to draw any firm conclusions, it seems that the early developments are encouraging. As much as 30% of the 1.8 trillion Euro post-COVID stimulus adopted by the European Union targets climate protection [28,29]. This most ambitious EU climate policy to date aims at spending around 0.55% of the EU GDP per year over 2021-2027, which would be consistent with the Paris Agreement objectives. Several individual European countries and some East Asian counties have also invested heavily in a green recovery [28,29]. In the United States, President Biden recently proposed the American Jobs Plan – an eight-year, 2 trillion USD investment package that is expected to allocate at least 1 trillion USD to the sectors related to clean energy, green infrastructure and environmental justice [30]. As noted optimistically in [31], even a small fraction of the total COVID-19 economic recovery packages announced by the governments around the globe could put the world on track to meet the Paris Agreement goals. However, the green stimulus packages currently implemented are short-term policies, while the successful transition to low-carbon technologies seems to require investment efforts in green technologies thaw would span several decades [32]. Moreover, green investment policies should be complemented by carbon pricing and subsidies for clean technology development and adoption. Therefore, while there is some promising early evidence, the long-run effect of COVID-19 on energy transition is surrounded by many uncertainties.

## 5. Conclusions

The COVD-19 pandemic has been a major shock that has drastically affected energy demand, $CO_2$ emissions and electricity systems around the world. The question of how emissions and renewable energy use will be affected in the long run by the pandemic is of paramount importance. In this paper, we attempt to illuminate this issue by looking at the empirical evidence from the past pandemic episodes. We used data on the six major pandemic crises that occurred between 1968 and 2016 and the local projection method that allows estimating the impulse responses of emissions and renewables' use to the pandemic shocks. Our results show that the previous pandemics led on average to the 3.4-3.7% fall in the $CO_2$ emissions in the short-run (1-2 years since the start of the pandemic). The effect is present only in the rich countries, as well as in countries with the highest pandemic death toll (where the effect disappears only after 8 years) and in countries that were hit by the pandemic during economic recessions. We found similar heterogeneity concerning the impact of past pandemics on the share of electricity generated from renewable sources. While on average we did not find a statistically significant impact, the effect is present for the OECD countries for which the past pandemics seem to increase the renewables' share of electricity by 1.9-2.3 percentage points within the five-year horizon. The previous pandemics have also increased the share of electricity generated from renewables in case of countries hit during bad economic times.

     Overall, our results show that the past pandemics have significantly reduced $CO_2$ emissions and accelerated the energy transition in the short- to medium run in the countries with the highest pandemic-related mortality, in the OECD countries, and in countries that experienced economic recession at the onset of the pandemic episodes. This suggests that pandemics create opportunities for green energy transitions. However, the sustainability of such a transition crucially depends on the determined and long-term government policy securing the required investment level in clean energy technologies during and after the pandemic episodes.

**Supplementary Appendix – for online publication only**

**Table A1.** List of pandemic events

| Announcement year | Event name | Affected countries (ISO codes) | Number of affected countries |
|---|---|---|---|
| 1968 | H3N2 Flu | ARG, AUS, CHL, DNK, FIN, FRA, GBR, GRC, HKG, ITA, JAM, JPN, NLD, NOR, PRT, SWE, USA, ZAF | 18 |
| 2003 | SARS | AUS, CAN, CHE, CHN, DEU, ESP, FRA, GBR, HKG, IDN, IND, IRL, ITA, KOR, KWT, MAC, MNG, MYS, NZL, PHL, ROU, RUS, SGP, SWE, THA, TWN, USA, VNM, ZAF | 29 |
| 2009 | H1N1 | AFG, AGO, ALB, AND, ARE, ARG, ARM, ATG, AUS, AUT, AZE, BDI, BEL, BGD, BGR, BHR, BHS, BIH, BLR, BLZ, BMU, BOL, BRA, BRB, BRN, BTN, BWA, CAN, CHE, CHL, CHN, CIV, CMR, COD, COG, COL, CPV, CRI, CUB, CYM, CYP, CZE, DEU, DJI, DMA, DNK, DOM, DZA, ECU, EGY, ESP, EST, ETH, FIN, FJI, FRA, FSM, GAB, GBR, GEO, GHA, GRC, GRD, GTM, GUY, HND, HRV, HTI, HUN, IDN, IND, IRL, IRN, IRQ, ISL, ISR, ITA, JAM, JOR, JPN, KAZ, KEN, KHM, KIR, KNA, KOR, KWT, LAO, LBN, LBY, LCA, LIE, LKA, LSO, LTU, LUX, LVA, MAR, MDA, MDG, MDV, MEX, MHL, MKD, MLI, MLT, MMR, MNE, MNG, MOZ, MUS, MWI, MYS, NAM, NGA, NIC, NLD, NOR, NPL, NRU, NZL, OMN, PAK, PAN, PER, PHL, PLW, PNG, POL, PRI, PRT, PRY, QAT, ROU, RUS, RWA, SAU, SDN, SGP, SLB, SLV, SOM, SRB, STP, SUR, SVK, SVN, SWE, SWZ, SYC, SYR, TCD, THA, TJK, TON, TTO, TUN, TUR, TUV, TZA, UGA, UKR, URY, USA, VCT, VEN, VNM, VUT, WSM, YEM, ZAF, ZMB, ZWE | 173 |
| 2012 | MERS | ARE, AUT, CHN, DEU, DZA, EGY, FRA, GBR, GRC, IRN, ITA, JOR, KOR, KWT, LBN, MYS, NLD, OMN, PHL, QAT, SAU, THA, TUN, TUR, USA, YEM | 26 |
| 2014 | Ebola | ESP, GBR, GIN, ITA, LBR, MLI, NGA, SEN, SLE, USA | 10 |
| 2016 | Zika | ABW, ARG, ATG, BHS, BLZ, BOL, BRA, BRB, CAN, CHL, COL, CRI, CUB, CYM, DMA, DOM, ECU, GRD, GTM, GUY, HND, HTI, JAM, KNA, LCA, NIC, PAN, PER, PRI, PRY, SLV, SUR, TCA, TTO, URY, USA, VCT, VEN | 38 |

*Source*: [14] and own construction.

**Table A2.** Summary statistics

|  | Mean | Standard deviation | Minimum | Maximum | $N$ |
|---|---|---|---|---|---|
| CO2 emissions per capita (CDIAC) | 4.43 | 8.10 | 0.00 | 168.43 | 11435 |
| CO2 emissions per capita (EDGAR) | 5.26 | 11.71 | 0.01 | 178.18 | 10100 |
| % of electricity generation from renewables | 32.83 | 34.26 | 0.00 | 100.00 | 7235 |
| GDP per capita (constant 2010 US$) | 11266.5 | 16721.6 | 132.1 | 141200.4 | 9014 |
| Trade (sum of exports and imports) (% of GDP) | 77.52 | 50.91 | 0.02 | 442.62 | 8149 |

*Source*: own construction. See section 2 in the main text for the description of data sources.

**Table A3.** Impact of past pandemics on CO2 emissions (CDIAC data, 1961-2017)

|  | k=0 | k=1 | k=2 | k=3 | k=4 | k=5 |
|---|---|---|---|---|---|---|
| $D_{i,t}$ | -0.015 | -0.034*** | -0.037*** | -0.004 | -0.022 | -0.006 |
|  | (0.012) | (0.011) | (0.014) | (0.017) | (0.017) | (0.015) |
| $D_{i,t-1}$ | -0.028*** | -0.043*** | -0.011 | -0.017 | 0.001 | 0.015 |
|  | (0.010) | (0.015) | (0.019) | (0.021) | (0.017) | (0.023) |
| $D_{i,t-2}$ | -0.033*** | -0.011 | -0.013 | 0.004 | 0.002 | -0.018 |
|  | (0.011) | (0.016) | (0.016) | (0.015) | (0.021) | (0.018) |
| $\Delta y_{i,t-1}$ | -0.142*** | 0.506*** | 0.626*** | 0.653*** | 0.684*** | 0.689*** |
|  | (0.023) | (0.023) | (0.029) | (0.027) | (0.026) | (0.026) |
| $\Delta y_{i,t-2}$ | -0.084*** | -0.343*** | -0.143*** | -0.045* | -0.010 | 0.034 |
|  | (0.018) | (0.021) | (0.035) | (0.023) | (0.025) | (0.021) |
| Real GDP per capita | -0.000*** | -0.000*** | -0.000*** | -0.000*** | -0.000*** | -0.000*** |
|  | (0.000) | (0.000) | (0.000) | (0.000) | (0.000) | (0.000) |
| Trade (% GDP) | 0.000 | 0.000 | 0.000 | -0.000 | -0.000 | -0.000 |
|  | (0.000) | (0.000) | (0.000) | (0.000) | (0.000) | (0.000) |
| Constant | 0.038 | 0.059 | 0.080* | 0.127** | 0.132*** | 0.114*** |
|  | (0.025) | (0.038) | (0.041) | (0.049) | (0.047) | (0.042) |
| Observations | 6823 | 6658 | 6494 | 6330 | 6166 | 6002 |
| Number of countries | 172 | 172 | 172 | 171 | 170 | 170 |
| Number of years | 58 | 57 | 56 | 55 | 54 | 53 |
| R-Square | 0.079 | 0.332 | 0.432 | 0.524 | 0.609 | 0.664 |

*Note*: results based on the OLS estimation of equation (1). See section 2 for the description of data sources. Standard errors in parentheses clustered at the country level. * $p < 0.1$, ** $p < 0.05$, *** $p < 0.01$. Country and time fixed effects are included but not reported.

**Table A4.** Impact of past pandemics on CO2 emissions (CDIAC data, 1961-2017) with high-severity pandemic dummy

|  | k=0 | k=1 | k=2 | k=3 | k=4 | k=5 |
|---|---|---|---|---|---|---|
| $D_{i,t}$ | -0.028* | -0.018 | -0.024 | -0.029* | -0.031* | -0.039** |
|  | (0.016) | (0.015) | (0.016) | (0.016) | (0.017) | (0.020) |
| $D_{i,t-1}$ | 0.001 | -0.001 | -0.006 | 0.012 | 0.003 | 0.036 |
|  | (0.014) | (0.016) | (0.019) | (0.023) | (0.020) | (0.023) |
| $D_{i,t-2}$ | -0.010 | -0.037** | -0.027* | -0.036** | -0.013 | -0.042** |
|  | (0.013) | (0.017) | (0.014) | (0.017) | (0.022) | (0.017) |
| $\Delta y_{i,t-1}$ | -0.141*** | 0.507*** | 0.627*** | 0.653*** | 0.684*** | 0.689*** |
|  | (0.023) | (0.023) | (0.029) | (0.028) | (0.026) | (0.026) |
| $\Delta y_{i,t-2}$ | -0.084*** | -0.343*** | -0.143*** | -0.046* | -0.010 | 0.033 |
|  | (0.018) | (0.021) | (0.035) | (0.023) | (0.025) | (0.021) |
| Real GDP per capita | -0.000*** | -0.000*** | -0.000*** | -0.000*** | -0.000*** | -0.000*** |
|  | (0.000) | (0.000) | (0.000) | (0.000) | (0.000) | (0.000) |
| Trade (% GDP) | 0.000 | 0.000 | 0.000 | -0.000 | -0.000 | -0.000 |
|  | (0.000) | (0.000) | (0.000) | (0.000) | (0.000) | (0.000) |
| Constant | 0.041* | 0.063 | 0.084** | 0.127** | 0.133*** | 0.113*** |
|  | (0.025) | (0.038) | (0.041) | (0.049) | (0.047) | (0.042) |
| Observations | 6823 | 6658 | 6494 | 6330 | 6166 | 6002 |
| Number of countries | 172 | 172 | 172 | 171 | 170 | 170 |
| Number of years | 58 | 57 | 56 | 55 | 54 | 53 |
| R-Square | 0.079 | 0.332 | 0.432 | 0.525 | 0.609 | 0.664 |

*Note*: results based on the OLS estimation of equation (1). See section 2 for the description of data sources. $D_{i,t}$ is a high-severity pandemic dummy (see section 2.1). Standard errors in parentheses clustered at the country level. * $p < 0.1$, ** $p < 0.05$, *** $p < 0.01$. Country and time fixed effects are included but not reported.

**Table A5.** Impact of past pandemics on CO2 emissions (CDIAC data, 1961-2017) with medium-severity pandemic dummy

|  | k=0 | k=1 | k=2 | k=3 | k=4 | k=5 |
|---|---|---|---|---|---|---|
| $D_{i,t}$ | 0.024 | -0.030 | -0.023 | 0.016 | -0.019 | -0.018 |
|  | (0.017) | (0.018) | (0.021) | (0.020) | (0.020) | (0.021) |
| $D_{i,t-1}$ | -0.036** | -0.030 | 0.002 | -0.036 | -0.032 | -0.018 |
|  | (0.017) | (0.022) | (0.020) | (0.024) | (0.023) | (0.023) |
| $D_{i,t-2}$ | -0.016 | 0.040** | 0.009 | 0.018 | 0.031 | 0.005 |
|  | (0.017) | (0.019) | (0.019) | (0.020) | (0.021) | (0.018) |
| $\Delta y_{i,t-1}$ | -0.141*** | 0.507*** | 0.626*** | 0.653*** | 0.684*** | 0.689*** |
|  | (0.023) | (0.023) | (0.029) | (0.027) | (0.026) | (0.026) |
| $\Delta y_{i,t-2}$ | -0.084*** | -0.342*** | -0.143*** | -0.046* | -0.010 | 0.034 |
|  | (0.018) | (0.021) | (0.035) | (0.023) | (0.025) | (0.021) |
| Real GDP per capita | -0.000*** | -0.000*** | -0.000*** | -0.000*** | -0.000*** | -0.000*** |
|  | (0.000) | (0.000) | (0.000) | (0.000) | (0.000) | (0.000) |
| Trade (% GDP) | 0.000 | 0.000 | 0.000 | -0.000 | -0.000 | -0.000 |
|  | (0.000) | (0.000) | (0.000) | (0.000) | (0.000) | (0.000) |
| Constant | 0.041 | 0.064* | 0.085** | 0.128*** | 0.132*** | 0.113*** |
|  | (0.025) | (0.038) | (0.041) | (0.049) | (0.046) | (0.042) |
| Observations | 6823 | 6658 | 6494 | 6330 | 6166 | 6002 |
| Number of countries | 172 | 172 | 172 | 171 | 170 | 170 |
| Number of years | 58 | 57 | 56 | 55 | 54 | 53 |
| R-Square | 0.079 | 0.332 | 0.432 | 0.525 | 0.609 | 0.664 |

*Note*: results based on the OLS estimation of equation (1). See section 2 for the description of data sources. $D_{i,t}$ is a medium-severity pandemic dummy (see section 2.1). Standard errors in parentheses clustered at the country level. * $p < 0.1$, ** $p < 0.05$, *** $p < 0.01$. Country and time fixed effects are included but not reported.

**Table A6.** Impact of past pandemics on CO2 emissions (CDIAC data, 1961-2017) with low-severity pandemic dummy

|  | k=0 | k=1 | k=2 | k=3 | k=4 | k=5 |
|---|---|---|---|---|---|---|
| $D_{i,t}$ | -0.010 | -0.011 | -0.008 | -0.009 | 0.008 | 0.039* |
|  | (0.011) | (0.013) | (0.017) | (0.019) | (0.021) | (0.020) |
| $D_{i,t-1}$ | -0.014 | -0.034** | -0.032 | -0.016 | 0.014 | -0.015 |
|  | (0.011) | (0.016) | (0.022) | (0.026) | (0.023) | (0.028) |
| $D_{i,t-2}$ | -0.024** | -0.033* | -0.010 | 0.017 | -0.016 | 0.011 |
|  | (0.012) | (0.019) | (0.018) | (0.019) | (0.022) | (0.021) |
| $\Delta y_{i,t-1}$ | -0.141*** | 0.506*** | 0.626*** | 0.653*** | 0.684*** | 0.689*** |
|  | (0.023) | (0.023) | (0.029) | (0.027) | (0.026) | (0.026) |
| $\Delta y_{i,t-2}$ | -0.084*** | -0.343*** | -0.143*** | -0.045* | -0.010 | 0.034 |
|  | (0.018) | (0.021) | (0.035) | (0.023) | (0.025) | (0.021) |
| Real GDP per capita | -0.000*** | -0.000*** | -0.000*** | -0.000*** | -0.000*** | -0.000*** |
|  | (0.000) | (0.000) | (0.000) | (0.000) | (0.000) | (0.000) |
| Trade (% GDP) | 0.000 | 0.000 | 0.000 | -0.000 | -0.000 | -0.000 |
|  | (0.000) | (0.000) | (0.000) | (0.000) | (0.000) | (0.000) |
| Constant | 0.045* | 0.068* | 0.087** | 0.128*** | 0.134*** | 0.113*** |
|  | (0.025) | (0.038) | (0.041) | (0.049) | (0.046) | (0.042) |
| Observations | 6823 | 6658 | 6494 | 6330 | 6166 | 6002 |
| Number of countries | 172 | 172 | 172 | 171 | 170 | 170 |
| Number of years | 58 | 57 | 56 | 55 | 54 | 53 |
| R-Square | 0.079 | 0.332 | 0.432 | 0.525 | 0.609 | 0.664 |

*Note*: results based on the OLS estimation of equation (1). See section 2 for the description of data sources. $D_{i,t}$ is a low-severity pandemic dummy (see section 2.1). Standard errors in parentheses clustered at the country level. * $p < 0.1$, ** $p < 0.05$, *** $p < 0.01$. Country and time fixed effects are included but not reported.

**Table A7.** Impact of past pandemics on $CO_2$ emissions (CDIAC data, 1961-2017) – the effects for the OECD versus non-OECD countries

|  | k=0 | k=1 | k=2 | k=3 | k=4 | k=5 |
|---|---|---|---|---|---|---|
| $D_{i,t}$ | -0.005 | -0.038*** | -0.028 | 0.003 | -0.023 | 0.009 |
|  | (0.014) | (0.013) | (0.018) | (0.020) | (0.020) | (0.018) |
| OECD dummy | 0.028 | 0.041 | 0.028 | 0.016 | 0.027 | 0.046* |
|  | (0.019) | (0.026) | (0.026) | (0.026) | (0.025) | (0.026) |
| $D_{i,t}$ *OECD dummy | -0.024** | 0.016 | -0.020 | -0.013 | 0.002 | -0.033** |
|  | (0.012) | (0.014) | (0.017) | (0.015) | (0.017) | (0.017) |
| $\Delta y_{i,t-1}$ | -0.141*** | 0.507*** | 0.626*** | 0.653*** | 0.684*** | 0.689*** |
|  | (0.023) | (0.023) | (0.029) | (0.027) | (0.026) | (0.026) |
| $\Delta y_{i,t-2}$ | -0.084*** | -0.343*** | -0.143*** | -0.045* | -0.010 | 0.034 |
|  | (0.018) | (0.021) | (0.035) | (0.023) | (0.025) | (0.021) |
| Real GDP per capita | -0.000** | -0.000*** | -0.000*** | -0.000*** | -0.000*** | -0.000*** |
|  | (0.000) | (0.000) | (0.000) | (0.000) | (0.000) | (0.000) |
| Trade (% GDP) | 0.000 | 0.000 | 0.000 | -0.000 | -0.000 | -0.000 |
|  | (0.000) | (0.000) | (0.000) | (0.000) | (0.000) | (0.000) |
| Constant | 0.041 | 0.064* | 0.082** | 0.128*** | 0.131*** | 0.113*** |
|  | (0.025) | (0.039) | (0.041) | (0.049) | (0.046) | (0.042) |
| AME of D at OECD | -0.029** | -0.022 | -0.048*** | -0.011 | -0.021 | -0.025 |
| Observations | 6823 | 6658 | 6494 | 6330 | 6166 | 6002 |
| Number of countries | 172 | 172 | 172 | 171 | 170 | 170 |
| Number of years | 58 | 57 | 56 | 55 | 54 | 53 |
| R-Square | 0.079 | 0.332 | 0.432 | 0.524 | 0.609 | 0.664 |

*Note*: results based on the OLS estimation of equation (2). AME is the average marginal effect of the pandemic dummy at the OECD countries. See section 2 for the description of data sources. Standard errors in parentheses clustered at the country level. * $p < 0.1$, ** $p < 0.05$, *** $p < 0.01$. Country and time fixed effects are included but not reported.

**Table A8.** Impact of past pandemics on $CO_2$ emissions (CDIAC data, 1961-2017) – the effects for lower-income versus higher-income countries

|  | k=0 | k=1 | k=2 | k=3 | k=4 | k=5 |
|---|---|---|---|---|---|---|
| $D_{i,t}$ | -0.017 | -0.039*** | -0.054*** | -0.007 | -0.035** | -0.021 |
|  | (0.013) | (0.012) | (0.015) | (0.017) | (0.016) | (0.016) |
| Lower-income countries dummy | -0.036** | -0.058*** | -0.075*** | -0.087*** | -0.067*** | -0.054** |
|  | (0.016) | (0.022) | (0.023) | (0.026) | (0.025) | (0.026) |
| $D_{i,t}$ * Lower-income countries dummy | 0.018 | 0.032* | 0.076*** | 0.017 | 0.072*** | 0.083*** |
|  | (0.019) | (0.019) | (0.021) | (0.027) | (0.027) | (0.029) |
| $\Delta y_{i,t-1}$ | -0.141*** | 0.506*** | 0.626*** | 0.653*** | 0.684*** | 0.688*** |
|  | (0.023) | (0.023) | (0.029) | (0.027) | (0.026) | (0.026) |
| $\Delta y_{i,t-2}$ | -0.084*** | -0.343*** | -0.143*** | -0.045* | -0.011 | 0.033 |
|  | (0.018) | (0.021) | (0.035) | (0.023) | (0.025) | (0.021) |
| Real GDP per capita | -0.000** | -0.000*** | -0.000*** | -0.000*** | -0.000*** | -0.000*** |
|  | (0.000) | (0.000) | (0.000) | (0.000) | (0.000) | (0.000) |
| Trade (% GDP) | 0.000 | 0.000 | 0.000 | -0.000 | -0.000 | -0.000 |
|  | (0.000) | (0.000) | (0.000) | (0.000) | (0.000) | (0.000) |
| Constant | 0.042* | 0.064* | 0.081** | 0.128*** | 0.130*** | 0.113*** |
|  | (0.025) | (0.038) | (0.041) | (0.049) | (0.046) | (0.042) |
| AME of D at non-OECD | 0.001 | -0.007 | 0.022 | 0.010 | 0.037 | 0.062** |
| Observations | 6823 | 6658 | 6494 | 6330 | 6166 | 6002 |
| Number of countries | 172 | 172 | 172 | 171 | 170 | 170 |
| Number of years | 58 | 57 | 56 | 55 | 54 | 53 |
| R-Square | 0.078 | 0.332 | 0.433 | 0.524 | 0.610 | 0.664 |

*Note*: results based on the OLS estimation of equation (2). AME is the average marginal effect of the pandemic dummy at the lower-income countries. Lower-income countries are low income and lower middle-income countries according to the World Bank classification. See section 2 for the description of data sources. Standard errors in parentheses clustered at the country level. * $p < 0.1$, ** $p < 0.05$, *** $p < 0.01$. Country and time fixed effects are included but not reported.

**Table A9.** Impact of past pandemics on CO2 emissions (CDIAC data, 1961-2017) – the role of macroeconomic conditions

|  | k=0 | k=1 | k=2 | k=3 | k=4 | k=5 |
|---|---|---|---|---|---|---|
| $(1 - F(z_{it})) * D_{i,t}$ | 0.040* | 0.001 | 0.027 | 0.007 | 0.009 | 0.045 |
|  | (0.021) | (0.021) | (0.021) | (0.025) | (0.030) | (0.030) |
| $F(z_{it}) * D_{i,t}$ | -0.063*** | -0.042** | -0.089*** | -0.028 | -0.044 | -0.064** |
|  | (0.021) | (0.019) | (0.023) | (0.029) | (0.031) | (0.030) |
| $\Delta y_{i,t-1}$ | -0.200*** | 0.482*** | 0.608*** | 0.641*** | 0.685*** | 0.708*** |
|  | (0.027) | (0.027) | (0.032) | (0.030) | (0.031) | (0.033) |
| $\Delta y_{i,t-2}$ | -0.089*** | -0.332*** | -0.104*** | -0.010 | 0.013 | 0.040* |
|  | (0.021) | (0.018) | (0.031) | (0.024) | (0.023) | (0.023) |
| $D_{i,t-1}$ | -0.015 | -0.037** | -0.016 | -0.018 | -0.005 | 0.003 |
|  | (0.010) | (0.015) | (0.018) | (0.020) | (0.017) | (0.022) |
| $D_{i,t-2}$ | -0.032*** | -0.018 | -0.020 | -0.007 | -0.004 | -0.023 |
|  | (0.012) | (0.018) | (0.017) | (0.016) | (0.021) | (0.020) |
| $g_{i,t-1}$ | 0.176 | -0.097 | -0.165 | 0.163 | -0.116 | -0.241* |
|  | (0.137) | (0.083) | (0.105) | (0.169) | (0.160) | (0.128) |
| $g_{i,t-2}$ | 0.021 | 0.254 | 0.509 | 0.257 | 0.202 | 0.211 |
|  | (0.082) | (0.203) | (0.335) | (0.276) | (0.145) | (0.219) |
| $F(z_{it-1})$ | -0.064** | -0.094*** | -0.036 | 0.022 | 0.001 | -0.018 |
|  | (0.026) | (0.022) | (0.024) | (0.032) | (0.034) | (0.031) |
| $F(z_{it-2})$ | -0.034* | 0.015 | 0.080 | 0.073 | 0.047 | 0.057 |
|  | (0.019) | (0.039) | (0.069) | (0.058) | (0.033) | (0.046) |
| Constant | 0.063* | 0.137*** | 0.053 | -0.016 | 0.035 | 0.063 |
|  | (0.033) | (0.034) | (0.044) | (0.064) | (0.052) | (0.046) |
| Observations | 7373 | 7193 | 7013 | 6834 | 6657 | 6479 |
| Number of countries | 180 | 180 | 180 | 179 | 179 | 179 |
| Number of years | 55 | 54 | 53 | 52 | 51 | 50 |
| R-Square | 0.106 | 0.322 | 0.427 | 0.527 | 0.615 | 0.675 |

*Note*: results based on the OLS estimation of equation (3). See section 2 for the description of data sources. Standard errors in parentheses clustered at the country level. * $p < 0.1$, ** $p < 0.05$, *** $p < 0.01$. Country and time fixed effects are included but not reported.

**Table A10.** Impact of past pandemics on the share of electricity generated from renewable sources (EIA data, 1980-2018)

|  | k=0 | k=1 | k=2 | k=3 | k=4 | k=5 |
|---|---|---|---|---|---|---|
| $D_{i,t}$ | -0.019 | 0.035 | 0.013 | 0.031 | 0.020 | 0.054 |
|  | (0.024) | (0.036) | (0.037) | (0.044) | (0.051) | (0.041) |
| $D_{i,t-1}$ | 0.047* | 0.021 | 0.049 | 0.044 | 0.098** | 0.064 |
|  | (0.027) | (0.037) | (0.037) | (0.039) | (0.039) | (0.047) |
| $D_{i,t-2}$ | 0.005 | 0.021 | 0.013 | 0.061 | 0.039 | -0.055 |
|  | (0.029) | (0.033) | (0.048) | (0.047) | (0.054) | (0.061) |
| $\Delta y_{i,t-1}$ | -0.129*** | 0.550*** | 0.646*** | 0.722*** | 0.723*** | 0.780*** |
|  | (0.038) | (0.034) | (0.055) | (0.048) | (0.054) | (0.046) |
| $\Delta y_{i,t-2}$ | -0.107*** | -0.384*** | -0.180*** | -0.111*** | -0.049 | -0.058* |
|  | (0.028) | (0.017) | (0.036) | (0.032) | (0.039) | (0.034) |
| Real GDP per capita | 0.000* | 0.000* | 0.000 | 0.000* | 0.000** | 0.000** |
|  | (0.000) | (0.000) | (0.000) | (0.000) | (0.000) | (0.000) |
| Trade (% GDP) | -0.000 | -0.000 | 0.000 | 0.000 | 0.000 | 0.000 |
|  | (0.000) | (0.000) | (0.000) | (0.000) | (0.000) | (0.000) |
| Constant | -0.003 | 0.026 | -0.009 | -0.100* | -0.196*** | -0.222*** |
|  | (0.047) | (0.056) | (0.065) | (0.058) | (0.063) | (0.072) |
| Observations | 4647 | 4489 | 4326 | 4163 | 4003 | 3845 |
| Number of countries | 174 | 174 | 173 | 168 | 164 | 162 |
| Number of years | 36 | 35 | 34 | 33 | 32 | 31 |
| R-Square | 0.123 | 0.411 | 0.502 | 0.593 | 0.654 | 0.711 |

*Note*: results based on the OLS estimation of equation (1). See section 2 for the description of data sources. Standard errors in parentheses clustered at the country level. * $p < 0.1$, ** $p < 0.05$, *** $p < 0.01$. Country and time fixed effects are included but not reported.

**Table A11.** Impact of past pandemics on the share of electricity generated from renewable sources (EIA data, 1980-2018) ) with high-severity pandemic dummy

|  | k=0 | k=1 | k=2 | k=3 | k=4 | k=5 |
|---|---|---|---|---|---|---|
| $D_{i,t}$ | 0.048 | 0.069 | 0.043 | 0.095* | 0.025 | 0.047 |
|  | (0.040) | (0.042) | (0.048) | (0.049) | (0.054) | (0.039) |
| $D_{i,t-1}$ | 0.038 | 0.018 | 0.053 | -0.017 | 0.001 | -0.028 |
|  | (0.043) | (0.051) | (0.055) | (0.044) | (0.047) | (0.041) |
| $D_{i,t-2}$ | 0.017 | 0.082 | 0.006 | 0.008 | -0.011 | -0.025 |
|  | (0.045) | (0.049) | (0.046) | (0.052) | (0.045) | (0.053) |
| $\Delta y_{i,t-1}$ | -0.129*** | 0.550*** | 0.646*** | 0.723*** | 0.724*** | 0.781*** |
|  | (0.038) | (0.034) | (0.055) | (0.048) | (0.054) | (0.046) |
| $\Delta y_{i,t-2}$ | -0.107*** | -0.385*** | -0.180*** | -0.112*** | -0.049 | -0.058* |
|  | (0.028) | (0.017) | (0.036) | (0.032) | (0.039) | (0.034) |
| Real GDP per capita | 0.000* | 0.000* | 0.000 | 0.000* | 0.000** | 0.000** |
|  | (0.000) | (0.000) | (0.000) | (0.000) | (0.000) | (0.000) |
| Trade (% GDP) | -0.000 | -0.000 | 0.000 | 0.000 | 0.000 | 0.000 |
|  | (0.000) | (0.000) | (0.000) | (0.000) | (0.000) | (0.000) |
| Constant | 0.001 | 0.033 | -0.010 | -0.109* | -0.209*** | -0.233*** |
|  | (0.047) | (0.055) | (0.064) | (0.057) | (0.061) | (0.070) |
| Observations | 4647 | 4489 | 4326 | 4163 | 4003 | 3845 |
| Number of countries | 174 | 174 | 173 | 168 | 164 | 162 |
| Number of years | 36 | 35 | 34 | 33 | 32 | 31 |
| R-Square | 0.123 | 0.412 | 0.502 | 0.594 | 0.654 | 0.711 |

*Note*: results based on the OLS estimation of equation (1). See section 2 for the description of data sources. $D_{i,t}$ is a high-severity pandemic dummy (see section 2.1). Standard errors in parentheses clustered at the country level. * $p < 0.1$, ** $p < 0.05$, *** $p < 0.01$. Country and time fixed effects are included but not reported.

**Table A12.** Impact of past pandemics on the share of electricity generated from renewable sources (EIA data, 1980-2018) with medium-severity pandemic dummy

|  | k=0 | k=1 | k=2 | k=3 | k=4 | k=5 |
|---|---|---|---|---|---|---|
| $D_{i,t}$ | -0.027 | -0.017 | -0.016 | -0.026 | 0.057 | 0.053 |
|  | (0.041) | (0.042) | (0.054) | (0.054) | (0.056) | (0.046) |
| $D_{i,t-1}$ | -0.004 | -0.025 | -0.036 | 0.052 | 0.058 | 0.078 |
|  | (0.037) | (0.041) | (0.042) | (0.047) | (0.059) | (0.057) |
| $D_{i,t-2}$ | -0.008 | -0.048 | 0.037 | 0.073 | 0.092 | 0.017 |
|  | (0.036) | (0.044) | (0.051) | (0.059) | (0.065) | (0.056) |
| $\Delta y_{i,t-1}$ | -0.129*** | 0.550*** | 0.647*** | 0.723*** | 0.723*** | 0.780*** |
|  | (0.038) | (0.034) | (0.055) | (0.048) | (0.054) | (0.046) |
| $\Delta y_{i,t-2}$ | -0.107*** | -0.384*** | -0.180*** | -0.111*** | -0.048 | -0.057* |
|  | (0.028) | (0.017) | (0.036) | (0.032) | (0.040) | (0.034) |
| Real GDP per capita | 0.000* | 0.000* | 0.000 | 0.000* | 0.000** | 0.000** |
|  | (0.000) | (0.000) | (0.000) | (0.000) | (0.000) | (0.000) |
| Trade (% GDP) | -0.000 | -0.000 | 0.000 | 0.000 | 0.000 | 0.000 |
|  | (0.000) | (0.000) | (0.000) | (0.000) | (0.000) | (0.000) |
| Constant | -0.002 | 0.027 | -0.012 | -0.105* | -0.197*** | -0.223*** |
|  | (0.047) | (0.055) | (0.065) | (0.057) | (0.063) | (0.072) |
| Observations | 4647 | 4489 | 4326 | 4163 | 4003 | 3845 |
| Number of countries | 174 | 174 | 173 | 168 | 164 | 162 |
| Number of years | 36 | 35 | 34 | 33 | 32 | 31 |
| R-Square | 0.122 | 0.411 | 0.502 | 0.594 | 0.654 | 0.711 |

*Note*: results based on the OLS estimation of equation (1). See section 2 for the description of data sources. . $D_{i,t}$ is a medium-severity pandemic dummy (see section 2.1). Standard errors in parentheses clustered at the country level. * $p < 0.1$, ** $p < 0.05$, *** $p < 0.01$. Country and time fixed effects are included but not reported.

**Table A13.** Impact of past pandemics on the share of electricity generated from renewable sources (EIA data, 1980-2018) with low-severity pandemic dummy

|  | k=0 | k=1 | k=2 | k=3 | k=4 | k=5 |
|---|---|---|---|---|---|---|
| $D_{i,t}$ | -0.034 | 0.009 | 0.006 | -0.011 | -0.065* | -0.030 |
|  | (0.022) | (0.029) | (0.042) | (0.037) | (0.036) | (0.034) |
| $D_{i,t-1}$ | 0.039 | 0.047 | 0.071 | 0.022 | 0.071 | 0.033 |
|  | (0.036) | (0.040) | (0.045) | (0.042) | (0.049) | (0.037) |
| $D_{i,t-2}$ | 0.012 | 0.019 | -0.025 | 0.009 | -0.027 | -0.047 |
|  | (0.035) | (0.040) | (0.048) | (0.043) | (0.065) | (0.052) |
| $\Delta y_{i,t-1}$ | -0.128*** | 0.550*** | 0.647*** | 0.723*** | 0.725*** | 0.781*** |
|  | (0.038) | (0.034) | (0.055) | (0.048) | (0.054) | (0.046) |
| $\Delta y_{i,t-2}$ | -0.107*** | -0.384*** | -0.180*** | -0.111*** | -0.050 | -0.058* |
|  | (0.028) | (0.017) | (0.036) | (0.032) | (0.039) | (0.034) |
| Real GDP per capita | 0.000* | 0.000* | 0.000 | 0.000* | 0.000** | 0.000** |
|  | (0.000) | (0.000) | (0.000) | (0.000) | (0.000) | (0.000) |
| Trade (% GDP) | -0.000 | -0.000 | 0.000 | 0.000 | 0.000 | 0.000 |
|  | (0.000) | (0.000) | (0.000) | (0.000) | (0.000) | (0.000) |
| Constant | -0.002 | 0.022 | -0.011 | -0.109* | -0.209*** | -0.230*** |
|  | (0.047) | (0.054) | (0.064) | (0.057) | (0.061) | (0.071) |
| Observations | 4647 | 4489 | 4326 | 4163 | 4003 | 3845 |
| Number of countries | 174 | 174 | 173 | 168 | 164 | 162 |
| Number of years | 36 | 35 | 34 | 33 | 32 | 31 |
| R-Square | 0.123 | 0.411 | 0.503 | 0.593 | 0.654 | 0.711 |

*Note*: results based on the OLS estimation of equation (1). See section 2 for the description of data sources. . $D_{i,t}$ is a low-severity pandemic dummy (see section 2.1). Standard errors in parentheses clustered at the country level. * $p < 0.1$, ** $p < 0.05$, *** $p < 0.01$. Country and time fixed effects are included but not reported.

**Table A14.** Impact of past pandemics on the share of electricity generated from renewable sources (EIA data, 1980-2018) – the effects for the OECD versus non-OECD countries

|  | k=0 | k=1 | k=2 | k=3 | k=4 | k=5 |
|---|---|---|---|---|---|---|
| $D_{i,t}$ | -0.042 | 0.019 | 0.003 | 0.014 | 0.007 | 0.036 |
|  | (0.030) | (0.044) | (0.046) | (0.053) | (0.063) | (0.047) |
| OECD dummy | -0.120** | -0.150** | -0.013 | 0.002 | 0.083 | 0.083 |
|  | (0.057) | (0.068) | (0.075) | (0.069) | (0.077) | (0.081) |
| $D_{i,t}$ *OECD dummy | 0.056 | 0.042 | 0.022 | 0.030 | 0.012 | 0.035 |
|  | (0.036) | (0.043) | (0.038) | (0.043) | (0.051) | (0.036) |
| $\Delta y_{i,t-1}$ | -0.129*** | 0.550*** | 0.646*** | 0.723*** | 0.724*** | 0.780*** |
|  | (0.038) | (0.034) | (0.055) | (0.048) | (0.054) | (0.046) |
| $\Delta y_{i,t-2}$ | -0.107*** | -0.384*** | -0.180*** | -0.111*** | -0.049 | -0.058* |
|  | (0.028) | (0.017) | (0.036) | (0.032) | (0.039) | (0.034) |
| Real GDP per capita | 0.000 | 0.000 | 0.000 | 0.000* | 0.000** | 0.000** |
|  | (0.000) | (0.000) | (0.000) | (0.000) | (0.000) | (0.000) |
| Trade (% GDP) | -0.000 | -0.000 | 0.000 | 0.000 | 0.000 | 0.000 |
|  | (0.000) | (0.000) | (0.000) | (0.000) | (0.000) | (0.000) |
| Constant | 0.008 | 0.032 | -0.009 | -0.106* | -0.207*** | -0.224*** |
|  | (0.048) | (0.056) | (0.064) | (0.057) | (0.061) | (0.071) |
| AME of $D$ at OECD | 0.014 | 0.061* | 0.025 | 0.044 | 0.020 | 0.071* |
| Observations | 4647 | 4489 | 4326 | 4163 | 4003 | 3845 |
| Number of countries | 174 | 174 | 173 | 168 | 164 | 162 |
| Number of years | 36 | 35 | 34 | 33 | 32 | 31 |
| R-Square | 0.123 | 0.411 | 0.502 | 0.593 | 0.654 | 0.711 |

*Note*: results based on the OLS estimation of equation (2). AME is the average marginal effect of the pandemic dummy at the OECD countries. See section 2 for the description of data sources. Standard errors in parentheses clustered at the country level. * $p < 0.1$, ** $p < 0.05$, *** $p < 0.01$. Country and time fixed effects are included but not reported.

**Table A15.** Impact of past pandemics on the share of electricity generated from renewable sources (EIA data, 1980-2018) – the effects for lower-income versus higher-income countries

|  | k=0 | k=1 | k=2 | k=3 | k=4 | k=5 |
|---|---|---|---|---|---|---|
| $D_{i,t}$ | -0.017 | 0.054 | 0.007 | 0.038 | 0.029 | 0.052 |
|  | (0.026) | (0.035) | (0.039) | (0.043) | (0.050) | (0.039) |
| Lower-income countries dummy | -0.039 | -0.043 | 0.035 | 0.112*** | 0.218*** | 0.233*** |
|  | (0.027) | (0.032) | (0.038) | (0.037) | (0.044) | (0.047) |
| $D_{i,t}$ * Lower-income countries dummy | -0.019 | -0.082 | 0.016 | -0.039 | -0.059 | -0.003 |
|  | (0.041) | (0.050) | (0.046) | (0.061) | (0.059) | (0.049) |
| $\Delta y_{i,t-1}$ | -0.129*** | 0.550*** | 0.646*** | 0.723*** | 0.725*** | 0.780*** |
|  | (0.038) | (0.034) | (0.055) | (0.048) | (0.054) | (0.046) |
| $\Delta y_{i,t-2}$ | -0.107*** | -0.384*** | -0.180*** | -0.112*** | -0.049 | -0.058* |
|  | (0.028) | (0.017) | (0.036) | (0.032) | (0.039) | (0.034) |
| Real GDP per capita | 0.000* | 0.000 | 0.000 | 0.000* | 0.000** | 0.000** |
|  | (0.000) | (0.000) | (0.000) | (0.000) | (0.000) | (0.000) |
| Trade (% GDP) | -0.000 | -0.000 | 0.000 | 0.000 | 0.000 | 0.000 |
|  | (0.000) | (0.000) | (0.000) | (0.000) | (0.000) | (0.000) |
| Constant | 0.001 | 0.025 | -0.012 | -0.106* | -0.206*** | -0.227*** |
|  | (0.047) | (0.055) | (0.064) | (0.057) | (0.062) | (0.071) |
| AME of $D$ at non-OECD | -0.036 | -0.028 | 0.023 | -0.002 | -0.031 | 0.049 |
| Observations | 4647 | 4489 | 4326 | 4163 | 4003 | 3845 |
| Number of countries | 174 | 174 | 173 | 168 | 164 | 162 |
| Number of years | 36 | 35 | 34 | 33 | 32 | 31 |
| R-Square | 0.122 | 0.411 | 0.502 | 0.593 | 0.654 | 0.711 |

*Note*: results based on the OLS estimation of equation (2). AME is the average marginal effect of the pandemic dummy at the non-OECD countries. See section 2 for the description of data sources. Standard errors in parentheses clustered at the country level. * $p < 0.1$, ** $p < 0.05$, *** $p < 0.01$. Country and time fixed effects are included but not reported.

**Table A16.** Impact of past pandemics on the share of electricity generated from renewable sources (EIA data, 1980-2018) – the role of macroeconomic conditions

|  | k=0 | k=1 | k=2 | k=3 | k=4 | k=5 |
|---|---|---|---|---|---|---|
| $(1 - F(z_{it})) * D_{i,t}$ | -0.071 | -0.050 | -0.007 | 0.013 | -0.011 | 0.005 |
|  | (0.044) | (0.073) | (0.059) | (0.059) | (0.066) | (0.063) |
| $F(z_{it}) * D_{i,t}$ | 0.023 | 0.100* | 0.035 | 0.041 | 0.048 | 0.116* |
|  | (0.040) | (0.061) | (0.065) | (0.060) | (0.080) | (0.059) |
| $\Delta y_{i,t-1}$ | -0.121*** | 0.539*** | 0.641*** | 0.728*** | 0.722*** | 0.777*** |
|  | (0.034) | (0.032) | (0.051) | (0.047) | (0.049) | (0.044) |
| $\Delta y_{i,t-2}$ | -0.107*** | -0.380*** | -0.194*** | -0.132*** | -0.068* | -0.075** |
|  | (0.027) | (0.017) | (0.033) | (0.032) | (0.036) | (0.034) |
| $D_{i,t-1}$ | 0.022 | 0.021 | 0.052 | 0.031 | 0.089** | 0.073 |
|  | (0.029) | (0.036) | (0.042) | (0.039) | (0.040) | (0.047) |
| $D_{i,t-2}$ | 0.003 | 0.039 | 0.018 | 0.068 | 0.063 | -0.017 |
|  | (0.028) | (0.040) | (0.047) | (0.046) | (0.054) | (0.064) |
| $g_{i,t-1}$ | 0.205* | -0.078 | -0.306 | -0.155 | -0.206 | -0.277 |
|  | (0.114) | (0.225) | (0.341) | (0.215) | (0.158) | (0.170) |
| $g_{i,t-2}$ | 0.008 | -0.256 | -0.184 | -0.181 | -0.208* | -0.323 |
|  | (0.111) | (0.225) | (0.112) | (0.111) | (0.113) | (0.209) |
| $F(z_{it-1})$ | 0.108*** | 0.054 | 0.006 | 0.047 | 0.016 | -0.030 |
|  | (0.034) | (0.052) | (0.072) | (0.053) | (0.046) | (0.049) |
| $F(z_{it-2})$ | 0.014 | -0.035 | -0.003 | -0.032 | -0.065* | -0.077 |
|  | (0.029) | (0.053) | (0.034) | (0.039) | (0.039) | (0.052) |
| Constant | -0.063 | 0.057 | 0.032 | -0.052 | -0.068 | -0.016 |
|  | (0.039) | (0.051) | (0.052) | (0.051) | (0.050) | (0.064) |
| Observations | 4920 | 4747 | 4571 | 4395 | 4223 | 4056 |
| Number of countries | 179 | 179 | 178 | 173 | 168 | 167 |
| Number of years | 36 | 35 | 34 | 33 | 32 | 31 |
| R-Square | 0.134 | 0.423 | 0.517 | 0.610 | 0.652 | 0.704 |

*Note*: results based on the OLS estimation of equation (3). See section 2 for the description of data sources. Standard errors in parentheses clustered at the country level. * $p < 0.1$, ** $p < 0.05$, *** $p < 0.01$. Country and time fixed effects are included but not reported.